# Structural origin of plasticity in strained high-entropy alloys

*Chi-Huan Tung,[†,‡] Guan-Rong Huang,[†] Zhitong Bai,[§] Yue Fan,[§] Wei-Ren Chen,[\*,†] and Shou-Yi Chang,[\*,‡]*

[†]Neutron Scattering Division, Oak Ridge National Laboratory, Oak Ridge, 37831, Tennessee, United States

[‡]Department of Materials Science and Engineering, National Tsing Hua University, Hsinchu 30013, Taiwan

[§]Department of Mechanical Engineering, University of Michigan, Ann Arbor, Michigan 48109, United States

**Abstract**

High-entropy alloys (HEAs) are solid solutions of multiple elements with equal atomic ratios which present an innovative pathway for de novo alloy engineering. While there exist extensive studies to ascertain the important structural aspects governing their mechanical behaviors, elucidating the underlying deformation mechanisms still remains a challenge. Using atomistic simulations, we probe the particle rearrangements in a yielding, model HEA system to understand the structural origin of its plasticity. We find the plastic deformation is initiated by irreversible topological fluctuations which tend to spatially localize in regions termed as soft spots which consist of particles actively participating in slow vibrational motions, an observation strikingly reminiscent of nonlinear glassy rheology. Due to the varying local elastic moduli resulting from the loss of compositional periodicity, these plastic responses exhibit significant spatial heterogeneity and are found to be inversely correlated with the distribution of local electronegativity. Further mechanical loading promotes the cooperativity among these local plastic events and triggers the formation of dislocation loops. As in strained crystalline solids, different dislocation loops can further merge together and propagate as the main carrier of large-scale plastic deformation. However, the energy barriers located at the spatial regions with higher local electronegativity severely hinders the motion of dislocations. By delineating the transient mechanical response in terms of atomic configuration, our computational findings shed new light on understanding the nature of plasticity of single-phase HEA.

One major enterprise in the field of metallurgy is to improve the mechanical strength of materials and to prevent materials from developing into multi-phase systems containing brittle intermetallic precipitates or incoherent phase boundaries which reduce their ductility.[1] In this pursuit, two remarkable papers, which were published respectively by Yeh[2] and Cantor[3] in 2004, demonstrated a strategy of creating single-phase alloys comprised of multiple elements with near-equal atomic ratios. A fascinating appeal of these so called high-entropy alloys (HEA) is that it promises an entirely new paradigm for engineering alloys with desired mechanical properties by circumventing the conflicting constraints of strength and ductility intrinsic to general metallic systems.[4,5] As is fully evident from the documented literature, in the past decade the field of HEA has undergone an explosive growth. Although an enormous range of desirable mechanical properties have already been demonstrated to be accessible in various HEA systems,[5-8] the fundamental mechanism governing their deformation behaviors remains to be explored. This challenge provides the main motivation of this computational work.

While the atomic lattice structure of HEA is built up by repetitive translation of the unit cell along the principal axes, the positional periodicity of each constituent element, presenting in most other crystalline solids, is no longer retained.[5] As a result, it renders an additional structural characteristic of disorder to HEA. In fact, existing x-ray and neutron diffraction studies,[9-11] especially those based on pair distribution function (PDF) analyses,[12] have demonstrated the dominant influence of this randomness in particle distribution on the structure of HEA. However, the fundamental question regarding how to address the mechanical properties of HEA from the perspective of its unique atomic arrangement has not been answered conclusively. Due to the lack of compositional periodicity, there exists some ambiguity as to whether or not the current theoretical framework for describing the structure–property correlation of ordered crystalline solids such as intermetallic compounds[13] including the Ni-Ti based shape memory alloys[14,15] are applicable in describing the mechanical behaviors of HEA.

From the peculiar crystal-glass hybrid character of HEA, we evaluate the mechanical response of strained HEA from a fundamental consideration of mechanistically addressing the configurational disorder embedded in a regular lattice. In the past decades the deformation mechanism of disordered materials has been the subject of intense study.[16] There exists an extensive amount of computational results aimed at elucidating the key microscopic features required to develop a theoretical description of their deformation behaviors.[16-24] Unlike the deformation of crystalline solids which can be described in terms of nucleation and propagation of dislocation,[25,26] the plastic events of disordered materials are often localized within the spatial regions where the constituent particles are involved in low-frequency vibrations.[27-30] Because of the irregular packing pattern of particles, the spatial distribution of this so-called "*soft spots*" in amorphous materials is found to be highly heterogeneous in space. In this setting an intriguing question naturally arises as whether the concept of soft spots, established for identifying the locations where the plastic instability in disordered solids takes place, can also be applied to HEA whose source of randomness comes from the irregular particle distribution over the set of ordered lattice points.

**Results**

**Spatial distribution of soft spots in stretched HEA.** To explore the existence of persistent soft spots in HEA, we computationally investigate the quinary CrMnFeCoNi alloy,[2] a model single-phase HEA system with a face-centered cubic (FCC) lattice, by means of atomistic simulations. Because severe lattice distortion has been demonstrated[3,5] to be a structural signature of HEA, our goal is to examine the connection between the structural deviation caused by substitutional impurities and the spatial distribution of soft spots. A defect-free FCC Ni crystal is also simulated as the reference system. To evaluate the intrinsic dissipation mechanism in HEA we calculate its vibrational modes based on an approach extended from that proposed by Mari and coworkers.[31] The atoms actively engaged in slow vibrational motions are identified by examining their participation fraction in 5% of the eigenmodes with the lowest frequencies. These values are further normalized by the mean value of the individual participation fraction. Examples of the calculated normalized participation fraction, which is defined as $\overline{P_i}$ for the $i^{th}$ atom in our system, are given in Figure 1.

As indicated by the calculated spatial distribution of $\overline{P_i}$, the particles participating in slow motions are seen to segregate together as those in amorphous materials. The spatial region characterized by $\overline{P_i} > 1$, which is marked by light blue color, can therefore be qualitatively regarded as the soft spots. To check the temporal fluctuation of $\overline{P_i}$ and how it influences the mechanical property, $\overline{P_i}$ obtained from averaging the displacement correlation matrix $\langle A \rangle_{ij}$ over short- and long-time interval are further calculated. Figures 1(a) and 1(b) give the short-time contour plots of $\overline{P_i}$ for Ni and CrMnFeCoNi respectively. The results are obtained via averaging $\langle A \rangle_{ij}$ over $1\ ps$. The heterogeneous nature of their instantaneous spatial distribution of $\overline{P_i}$ is clearly revealed. We further calculate $\overline{P_i}$ from averaging $\langle A \rangle_{ij}$ over $1\ ns$ and present the results in Figures 1(c) to 1(d). In comparison to the instantaneous snapshots, this statistical averaging is found to smear the fluctuation in the spatial distribution of $\overline{P_i}$ for Ni crystal. This observation is consistent with the expectation from the atomic settings implemented in our simulations: In a defect-free FCC Ni crystal, the translational symmetry is preserved and thus the local topology for each atom is essentially identical. It can therefore be inferred that the low-frequency vibration in a perfect crystal is a short-lived stochastic event which is driven thermodynamically. Moreover, its transient nature is reflected by the reducing heterogeneity revealed by the comparison of Figures 1(a) and 1(c).

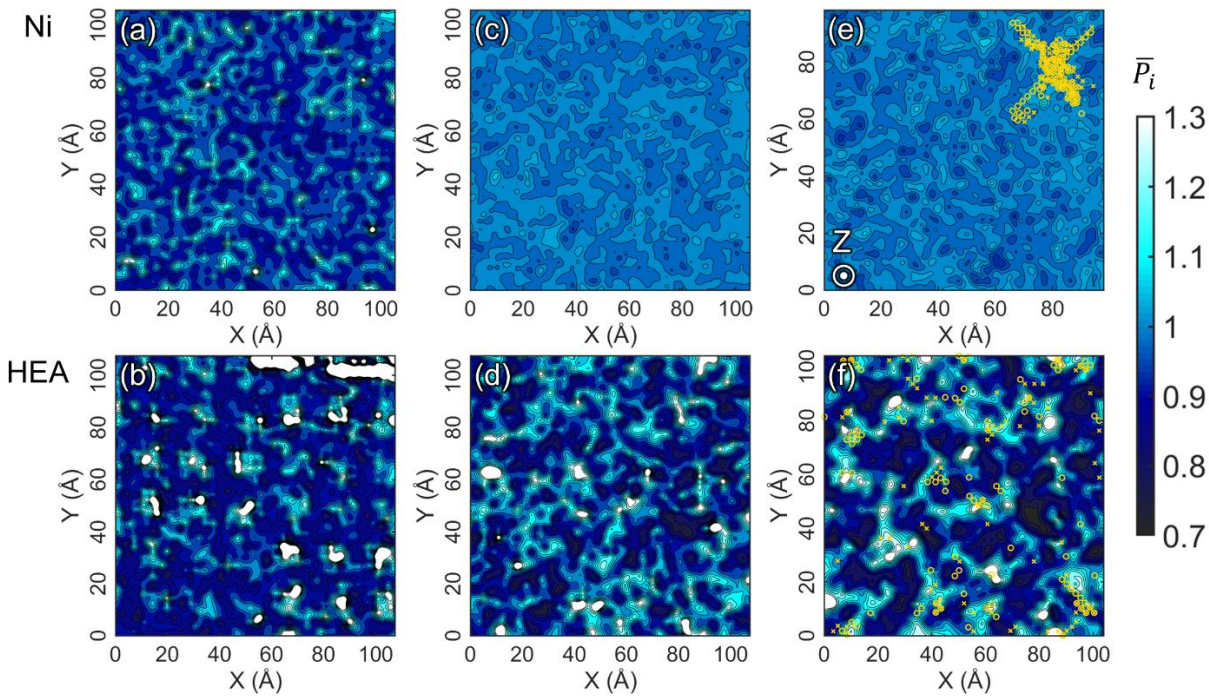

**Fig. 1** Short-time contour plots of normalized atomic participation fraction of slow vibrational dynamics $\overline{P}_i$ in a two-layer slice for (a) defect-free FCC nickel and (b) CrMnFeCoNi HEA. The normal vector of the plane is parallel to the direction of applied tensile stress. In both cases the heterogeneity of their spatial distribution is observed. The long-time averaged counterparts for nickel and CrMnFeCoNi are given in panels (c) and (d) respectively. The comparison reveals that the heterogeneity in the spatial distribution of $\overline{P}_i$ is considerably smeared and indicates the difference in the statistical nature of slow vibrations for both simulated systems. The yellow symbols in Panels (e) and (f) indicate the locations of nonaffine particle rearrangements in nickel and CrMnFeCoNi subjected to applied tensile forces, where the circles display particles with notable non-affine parameter $D_i^2$ and crosses for local strain $\gamma_l$ respectively (Methods). As in strained amorphous solids, these local plastic events in stretched HEA are seen to localized at spatial regions colored light blue where the constituent atoms actively participate in slow vibrational motions.

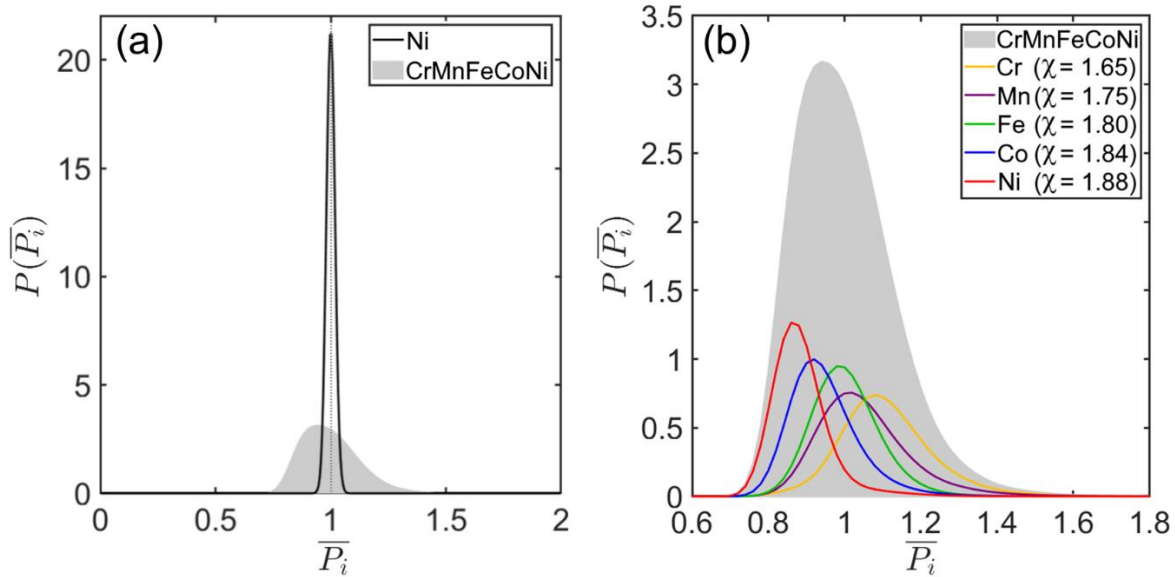

**Fig. 2** (a) Population density of participation fraction $\overline{P}_i$, $P(\overline{P}_i)$, for nickel and CrMnFeCoNi. The narrow, symmetric distribution of $P_{\text{Ni}}(\overline{P}_i)$ for the reference nickel crystal can be satisfactorily described by a Gaussian function. On the contrary, the shape of $P_{\text{CrMnFeCoNi}}(\overline{P}_i)$ is seen to be more skewed, indicating an increase in the low-frequency vibrational degree of freedom. (b) The individual contribution of constituent elements to $P_{\text{CrMnFeCoNi}}(\overline{P}_i)$. Atoms with lower electronegativity are seen to exhibit higher participation in low-frequency collective excitation.

**The negative correlation between soft spots and local electronegativity.** The transition in the intrinsic characteristics of clustering phenomenon of $\overline{P}_i$ from a stochastic event for a Ni crystal to a deterministic process for HEA can be appreciated from inspecting Figures 1(a) to 1(d). Moreover, not only the spatial distribution of $\overline{P}_i$ in CrMnFeCoNi alloy becomes more persistent, it also appears to be more heterogeneous. The physical origin of this observation is worth scrutinizing: A recent density-functional theory (DFT)[32] calculation of HEA has addressed the connection of the charge transfer between the neighboring elements and the local stress field. Many key properties of HEA are further correlated with the difference in electronegativity $\chi$ among the constituent elements. The implication of this electronic structure calculation has brought out an intriguing possibility that a quantitative description regarding the complicated information of low-frequency collective excitations in HEA can also be distilled and couched in terms of $\chi$.

To verify this hypothesis first we calculate the population density of $\bar{P}_l$, $P(\bar{P}_l)$, and give the results in Figure 2(a): For defect-free FCC Ni crystal, $\bar{P}_l$ is seen to narrowly distribute around the mean value of $\bar{P}_l = 1$. Moreover, the distribution can be satisfactorily described by a Gaussian function with an explicit expression of $P_{\text{Ni}}(\bar{P}_l) = \frac{1}{0.016\sqrt{2\pi}} exp\left[-\frac{(\bar{P}_l-1)^2}{2\times 0.016^2}\right]$. Unlike the Gaussian lineshape for Ni, $P(\bar{P}_l)$ for CrMnFeCoNi alloy is discernibly skewed. $\bar{P}_l$ is found to distribute within the range from 0.6 to 1.8, reflecting the significant increase in the low-frequency vibrational degree of freedom. We further decompose $P_{\text{CrMnFeCoNi}}(\bar{P}_l)$ into the individual contribution of each constituent element. The result given in Figure 2(b) reveals a general trend that a constituent element with lower electronegativity exhibits higher participation in low-frequency collective excitation. Existing studies of amorphous materials have demonstrated that their bulk mechanical properties show a strong dependence on local atomic packing.[33] In this context, because HEA is also characterized by structural inhomogeneity at atomic level, to address its threshold of plastic instability it becomes clear that the ensemble average of any relevant physical quantity does not necessarily provide the most insightful information to understand the diverse phenomenon of local topological rearrangements. Instead, we argue the local average provides an intuitively more direct reflection at the atomistic-level mechanisms. In this study the excess local average of a given physical quantity $\alpha$ around a point of interest $i$ is obtained by summing $\alpha$ for all particles presenting in a spatial range defined by a Gaussian field with a standard deviation $\sigma$ set to be the nearest neighbor distance defined by the local PDF and subtracted by the global average of $\alpha$. The explicit expression is $\langle \alpha_i \rangle = \frac{\sum_j exp\left(-\frac{r_{ij}^2}{2\sigma^2}\right)\alpha_j}{\sum_j exp\left(-\frac{r_{ij}^2}{2\sigma^2}\right)} - \frac{1}{N}\sum_{j=1}^{N} \alpha_j$, where $r_{ij}$ is the distance between particles $i$ and $j$ and N is the particle number.

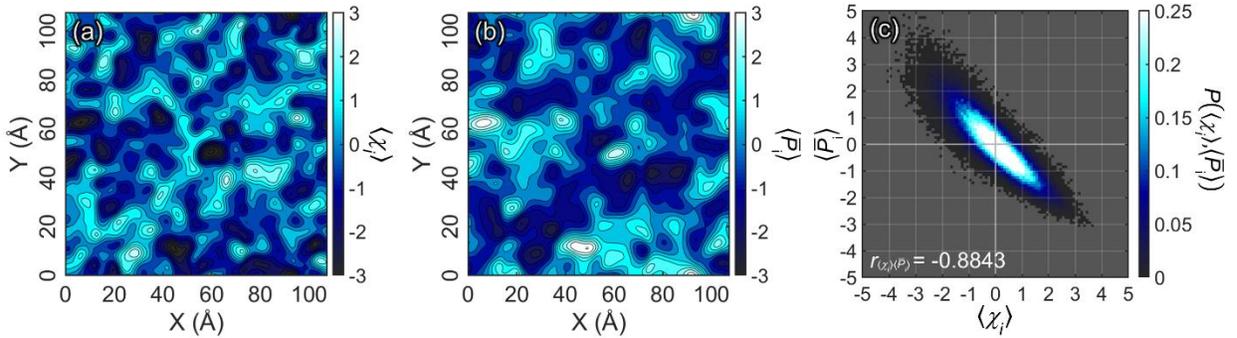

**Fig. 3** Panels (a) and (b) give the maps of averaged local electronegativity $\langle \chi_i \rangle$ and averaged local atomic participation fraction $\langle \overline{P}_\iota \rangle$, respectively, in CrMnFeCoNi. A negative correlation can be identified by visual inspection of the color schemes. A calculation of Pearson correlation coefficient $r_{\langle \chi_i \rangle \langle \overline{P}_\iota \rangle}$ is conducted to quantify this qualitative observation. A strong negative correlation is revealed by the calculated value of $r_{\langle \chi_i \rangle \langle \overline{P}_\iota \rangle} = -0.8843$.

The spatial distributions of standardized $\langle \chi_i \rangle$ and $\langle \overline{P}_\iota \rangle$ are presented in Figures 3. From the shapes, patterns, and the spacing of the contour lines with equal numerical values, it is seen that the spatial regions characterized by higher $\langle \chi_i \rangle$ in Figure 3(a) essentially overlap those with lower $\langle \overline{P}_\iota \rangle$ in Figure 3(b). This negative correlation is further illustrated by the two-dimensional density plot presented in Figure 3(c). The Pearson correlation coefficient $r_{\langle \chi_i \rangle \langle \overline{P}_\iota \rangle}$ is calculated to quantitatively measure the linear correlation between $\langle \chi_i \rangle$ and $\langle \overline{P_i} \rangle$ and is found to be -0.8843. This result clearly shows that particles present in the spatial regions with lower $\langle \chi_i \rangle$ exhibit more active participation to the low-frequency vibrations. Among the constituent elements of our simulated HEA system, chromium and manganese are characterized by lowest $\chi$, which are 1.65 and 1.75 respectively in Allen's scale[34]. We therefore conclude that the Cr or Mn-rich sites are the spatial origin of soft spots in CrMnFeCoNi alloy. An important phenomenon that is closely related to Figure 3 is the influence of $\langle \chi_i \rangle$ on plastic instability in amorphous alloys. Judging from the related studies,[33,35] it is inferred that the spatial regions characterized by both positive and negative anomalies of $\langle \chi_i \rangle$ are mechanically unstable and more likely to respectively accommodate topological transformation of expansion and compression when mechanical loading drives the system pass the local instability threshold. The negative correlation between $\langle \chi_i \rangle$ and $\langle \overline{P_i} \rangle$ given in Figure 3(c) suggests that this stability criterion of amorphous alloys is not valid in HEAs. Instead of exhibiting positive density fluctuation due to the excess atomic-level compressive pressure expected from the related DFT calculation,[11] spatial regions with higher $\langle \chi_i \rangle$ remains mechanically stable during the deformation process. It is our conjecture that the origin of this observation is the anharmonicity of interaction potential among the constituent elements: Given the same amount of strain, the energy cost and restoring force for compressive deformation is considerably larger than those of tensile deformation taking place in the soft spots characterized by lower $\langle \chi_i \rangle$.

**Nonaffine deformation and soft spots.** We further apply a volume preserved tensile strain with a strain rate $\dot{\gamma}_T$ of $5 \times 10^8 \, s^{-1}$ to examine the local yielding events in the mechanically driven metallic systems. When subjected to small elongation, both systems exhibit affine motions. Approaching the yield point, the local deformation is no longer uniform. The displacements of particles become progressively nonaffine and consequently the local strain is heterogeneous. We calculate the spatial distributions of long-time averaged $\overline{P_l}$ just beyond the yield point and give the results in Figures 1(e) to 1(f). There exist several approaches to evaluate the nonaffinity in strained materials. One way to define the nonaffine displacement is by comparing the particle displacement with the ideal affine condition and provide a scalar description of local nonaffinity from the deviation. Based on this mathematical framework two methods developed by Langer[18] and Li[36] respectively are adopted in this work and the results are marked by yellow symbols in Figures 1(e) and 1(f). Evidenced by the formation of intersecting multiplanar slips belonging to different slip systems in Figure 1(e), the defect-free FCC Ni crystal is found to yield abruptly and followed by dislocation induced plasticity after yielding. This observation is consistent with the earlier report by Zepeda-Ruiz and coworkers[37]. No discernible correlation between the locations of the plastic rearrangement and the spatial distribution of $\overline{P_l}$ is found. On the contrary, for the results of CrMnFeCoNi HEA given in Figures 1(f), the spatial distribution of $\overline{P_l}$ is discernibly more heterogeneous than that in FCC Ni crystal. Moreover, the identified nonaffine displacements are found to be essentially distributed in the light blue regions of the spatial distribution of $\overline{P_l}$.

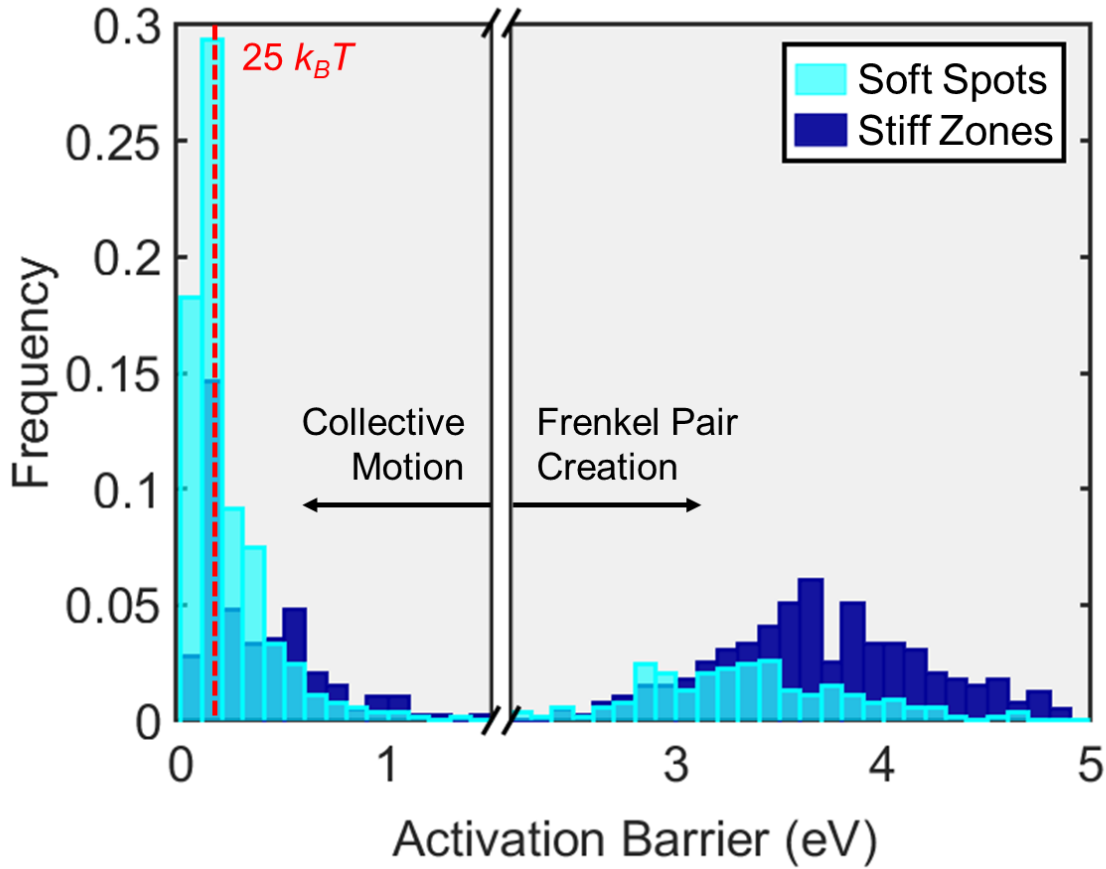

**Fig. 4** Distributions of activation barriers for atomic rearrangements in soft spots (light blue) and stiff zones (dark blue), respectively. Evidenced by the related studies[46], the activation energy barrier for dislocation induced plasticity is around $25\ k_BT$. Such activation spectra are obtained by sampling the system's underlying potential energy landscape prior to the yield point using the activation–relaxation technique (ART) algorithm[41]. The activation barriers for atomic rearrangements in soft spots, in particular for the non-affine collective motion, are statistically much smaller in comparison to those taking place in stiff zones such as Frenkel pairs.

**Difference in energy barriers of nonaffine deformation between soft spots and stiff zones.** To further characterize the differences between the soft spots, namely the light blue regions in Figure 1, and the relatively stiff spots colored by dark blue, we probe their underlying potential energy landscape (PEL) structures, because it is known that emergent deformation units, such as atomic rearrangements, in a condensed matter system should correspond to the hopping between contiguous local minima in the PEL. In the present study the PEL structures are explored via activation relaxation technique (ART), which is known capable of providing the key connections between neighboring metastable states in the PEL without invoking empirical assumptions[38-40]. Specifically, small random displacement perturbations are introduced in a location-specific manner to both the soft and stiff regions slightly before the yield point. The responses of the system are then characterized by the ART search algorithm[41], which enables the identifications of deformation mechanisms and their corresponding activation energy spectra[42-45].

As illustrated in Figure 4, two mechanisms are present for atomic rearrangements, namely a low-barrier collective motion mode (e.g. vortex displacements discussed below, see also in Supplementary Materials for details), and a high-barrier Frenkel pair mode, respectively. In stark contrast to the stiff zone, the activation barriers in the soft zone are statistically smaller. The activation barrier for the plastic events induced by the migration of pre-exist dislocation in crystal, which is typically around $25\ k_B T$, is also indicated by the dashed line in Figure 4.[46] More importantly, the low-barrier collective motion mode in the soft spots is significantly more enhanced, corroborating well with the parallel von Mises strain and nonaffine displacement analyses in MD simulations (i.e. yellow marks in Figure 1(e)-(f)). Such validations from independent PEL calculations therefore strengthen the key hypothesis of our study, indicating that the concept of soft spots can be extended to facilitate the mechanical study of highly distorted crystalline systems such as HEA, which will warrant future studies.

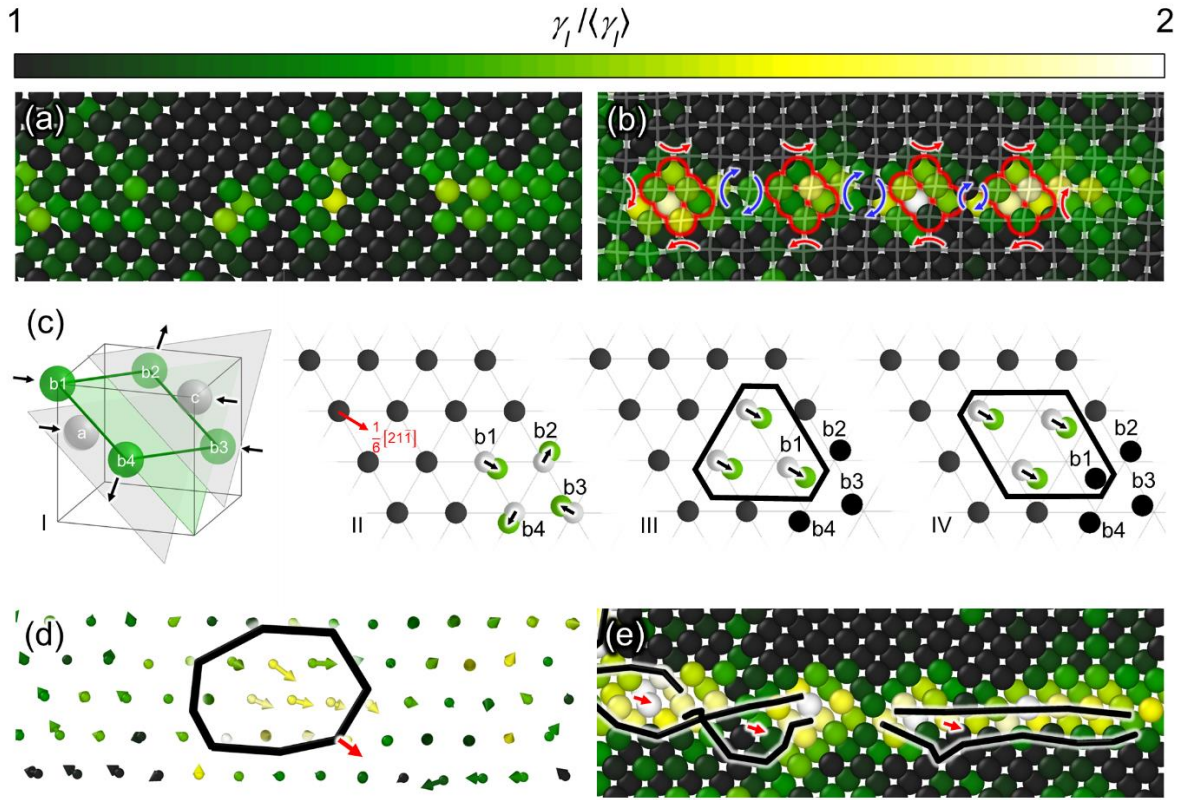

**Fig. 5** Panel (a) gives an instantaneous spatial distribution of normalized local strain, $\gamma_l/\langle\gamma_l\rangle$ in the early loading stage. $\langle\gamma_l\rangle$ is the averaged atomic strain. Formation of localized deformation patches (LDPs) on $(001)$ plane perpendicular to the tensile direction is revealed by the color scheme. As illustrated in Panel (b), further increase of mechanical loading leads to the formation of vortex units (shown in red) which promote the cooperativity among the isolated LDPs. Panel (c) gives an example demonstrating how a dislocation loop can be formed and developed on $(111)$ plane from the collective motions of particles of a LDP. The direction of Burgers vector is given by the red arrow. As indicated in Panel (d), a discontinuity in the particle displacement is clearly defined by the hybrid dislocation loop developed from LDPs. The snapshot of local strain field given in Panel (e) demonstrates the interaction between two dislocation loops. It can be visualized that the strained soft spots can be segments or termini of dislocation loops.

**The influence of soft spots on the formation of dislocation loops.** We can now readily examine the influence of soft spots on the deformation mechanism of HEA at the microscale. It is instructive to evaluate the structural response of HEA to applied mechanical forces and such results are presented in Figure 5. In this figure we examine the atomic configuration of a soft spot region in CrMnFeCoNi HEA during a plastic event in terms of local strain $\gamma_l$. Under uniaxial loading, strain begins to rise. As illustrated by the sequential color scheme given in Figure 5(a), during the early loading stage localized deformation patch (LDP) begin to develop on the (001) plane perpendicular to the tensile direction. The origin of this observed strain localization is attributed to the fluctuation of local elastic moduli[47,48]. Our trajectory analysis shows that on average each individual patch consists of 12-14 atoms. This size appears to be considerably smaller than that of a shear transformation zone (STZ) in metallic glasses,[49] which is estimated from around twenty[50-52] to several hundred atomic volume.[53,54]

In the vicinity of yield point, a characteristic development of the local strain field is further revealed in Figure 5(b): More STZ-like LDPs are created by the continuous loading of mechanical energy. Moreover, similar to the topological evolution of strained metallic glasses near the instability threshold,[55] formation of localized vortex units which essentially conduct affine rotation is also observed in our simulation. The spatial arrangement of these LDPs and vortex units are compelled to follow certain dispositions in terms of location and symmetry. As a result, they appear in an alternate fashion such that each LDP is neighbored by two vortex units and vice-versa in order to maintain force balance at atomic level.

As indicated by the sketch of the collective particle motion in Figure 5(b), the presence of these vortex units promotes the cooperativity among the strained patches. In metallic glasses, the mechanical failure is known to be controlled by the interaction among these localized strained patches, viz. the phenomenon of shear banding. While the nucleation process of strain localization in HEA is seen to be similar to that of metallic glasses in terms of configurational variation, in the later stage of plastic yielding one spectacular difference is revealed by our trajectory analysis as illustrated in Figure 5(c): Panel I gives the atomic configuration of the nearest-neighbor structure engaging in a local deformation event. For the sake of simplicity, particles are placed at the FCC lattice points and the structural distortion due to the multielement effect is not incorporated. In this example, the tensile stress along the [001] direction induces a local shear field. At the onset of topological instability, particles $b1$ and its second-nearest neighbor $c$ which belong to two adjacent (111) planes, as well as $a$ and $b3$, are pushed towards each other by the local compressive stress. Meanwhile, particles $b2$ and its first-nearest neighbor $b4$ located at the same (111) plane move further away along the direction of shear stress. In Panels II-IV we inspect the evolution of particle displacement of an (111) plane which intersects the (001) plane in a line reflected by the highly strained linear region given in Figure 5(b). The competition of compression and dilation effects is revealed by the collective movement $b1$, $b2$, $b3$ and $b4$ on the shear flow-vorticity plane. The compressional deformation alone $\overrightarrow{b1b3}$ triggers a sequential collective particle movement presented in Panel (III) to compensate the loss of line density. From the map of particle displacement given in Figure 5(d), a dislocation loop, within which the particles are highly strained as indicated by the color scheme, can be visualized from the discontinuity of displacement field. Trajectory analysis based on the dislocation extraction algorithm (DXA)[56] algorithm demonstrates a pristine hybrid dislocation loop, characterized by the Burgers vector $b = \frac{1}{6}[21\bar{1}]$, is created on this (111) plane. This Burgers vector analysis shows that further increase of strain leads to the expansion of dislocation loop as illustrated in Panel (IV).

Although these observations demonstrate that the dissipative motion of dislocations remains an important mechanism responsible for releasing the loaded mechanical energy in strained HEA, it is instructive to point out a mechanistic subtlety in the spatial array of dislocation which again manifests the inherent nature of disorder in HEA: Because dislocations define the line of demarcation of an area over which a crystalline solid is strained, it has been mathematically demonstrated that they cannot end inside the crystalline body.[25] However, due to the irregularly distorted lattice, the propensity of dislocations to propagate inside HEA is not restricted by this conservation law of dislocations for conventional crystalline solids. As demonstrated in Figure 5(e), the highly strained LDPs can also act as a source or sink for dislocations in HEA. Depending on the details of atomic packing, different dislocation loops could coexist at different locations with different local stress fields in a strained HEA crystal. As in conventional crystalline solids, these pristine dislocation loops in HEA can further merge together to form a larger loop and propagate across the HEA crystal via the certain slip system, which is energetically favorable.

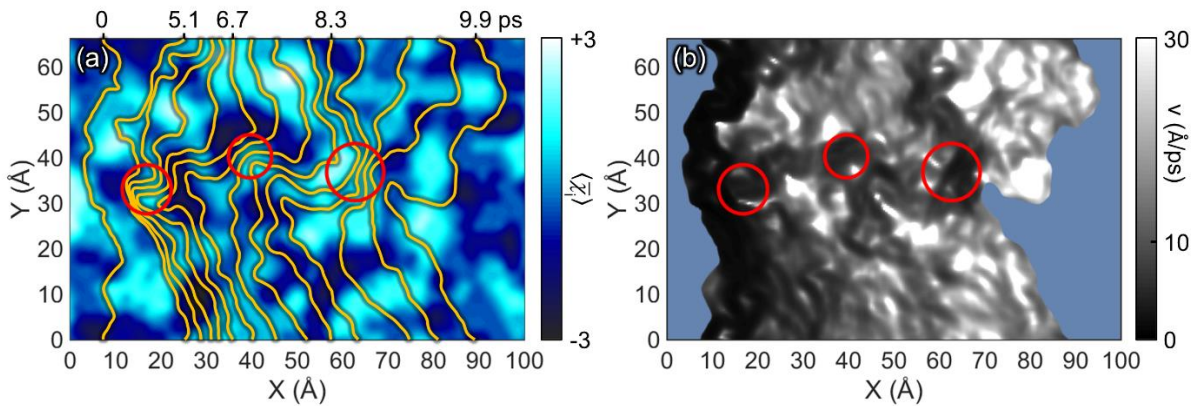

**Fig. 6** (a) The positions of a dissipative dislocation on the glide (111) plane at different time. The background is the spatial distribution of $\langle \chi_i \rangle$. At any given time the dislocation is characterized by a wavy shape. (b) The spatial distribution of the dislocations' instantaneous velocity $v$. The average velocity is estimated to be $10\ \text{Å}/ps$. By inspecting the denoted areas denoted with red circles in Panels (a) and (b), a connection between $\langle \chi_i \rangle$ and segmental dissipativity of dislocation can be established.

**The connection between $\langle \chi_i \rangle$ and dislocation mobility.** To further explore the underlying atomistic mechanisms governing the complex plastic deformation behavior in strained HEAs, we investigate the dislocation mobility and give the results in Figure 6. Panel (a) presents the position of a dislocation as a function of time on the glide (111) plane. Similar to the previous reports,[57-59] in any given instance a linear dislocation of irregular shape is observed. One can argue that the shape of a dislocation in HEA is determined by the thermodynamic tradeoff between two counteracting effect, the elastic strain energy determined by the contour length of a dislocation, and the thermal energy required for lifting a dislocation collectively over the energy barriers along the line. While less energy is required for triggering a non-affine atomic rearrangement at soft spots as indicated by Figure 4, the penalty of elastic strain energy prevents a dislocation from fully developing along the soft spots. Therefore, the dislocation tends to become wavy as illustrated in Figure 6(a). Due to the fluctuations in local moduli, the dislocation motion is driven by local stresses which are also highly heterogeneous. As a result, the complex dissipation behavior of a dislocation in stretched HEA is reflected by the spatial distribution of its instantaneous velocity $v$ given in Figure 6(b). From the color scheme, one can identify the darker spatial regions as the locations where the dislocation motion is impeded. While in dilute solid solutions it is well recognized that the pinning of dislocations is mainly caused by the solute atoms, to date no obvious analogy exists to locate the intrinsic barriers for dislocation motion in HEAs. The result given in Figure 3 raises a possibility that the averaged local electronegativity $\langle \chi_i \rangle$ can be used as such criterion of structural distinction. Indeed, by inspecting the distribution of $\langle \chi_i \rangle$ and $v$ of three denoted areas given in Figures 6(a) and 6(b), less dissipation is observed within the spatial region characterized by higher $\langle \chi_i \rangle$. This observation demonstrates the dislocation motion is indeed impeded by these stiff zones in HEAs.

**Discussion**

In summary, we have computationally investigated the deformation behavior of CrMnFeCoNi HEA. First the constituent atoms which actively engage in low frequency vibration motions are found to aggregate sporadically in space. We show this clustering phenomenon, which is defined as soft spots, is strongly correlated with the spatial distribution of local electronegativity. As in strained amorphous solids, plastic deformation in stretched HEA is found to be initiated by isolated topological rearrangements within soft spots due to lower activation energy. These local plastic events can be thermally activated and collectively form dislocation loops as in crystalline solids. The dislocation loops can further coalesce and lead to large-scale plastic deformation. We find that while the motion of dislocations can still be described in term of a certain slip system, it is hindered by the local energy barriers which governs the stress strain relationship microscopically. Results of this computational study thus suggest that it is the heterogeneity in local modulus caused by the random distribution of particles on crystalline lattice which renders the increase in the mechanical strength of HEA without compromising the ductility.

In addition to being a computational step to address the structural origin responsible for the plasticity of HEA, the present work has indicated that the structural characterization of soft spots might be of interest in its own right. Because our simulation has demonstrated that dislocation lines inside HEA can either be created or end at soft spots, it is reasonable to infer that soft spots should be characterized by some specific topological and geometric features, in terms of atomic packing, spatial symmetry and compositional dependence, which are distinguishably different from those of elastic matrix and thus make the nucleation and termination of dislocations possible.

One current research focus of HEA is to experimentally explore the existence of chemical short-range ordering (CSRO),[3,58-65] referring to the aggregation of certain constituent elements within a few atomic spacings, and evaluate how it influences the mechanical properties of HEA. While it is not possible for us to directly address its effect on the plasticity of HEA due to the limitation of the used force field, the connection between this loss of local atomic random dispersion and soft spots in HEA can still be intuitively inferred from the result of this computational study: Existing studies suggest that in HEA this clustering phenomenon only takes place sporadically in space,[58-61] which in turn indicates that CSRO is not characterized by any long-range translational symmetry. Therefore, it can be inferred that the existence of CSRO only changes the spatial distribution and lifetime of soft spots in strained HEA in a quantitative manner and the qualitative features of heterogeneity and intermittency are expected to remain intact. Investigating the influence of CSRO on the plasticity of HEA, from the perspectives of soft spots and local electronegativity should be an interesting area ripe for future computational research.

**Methods**

**Atomistic molecular dynamics (MD) simulation.** The initial configuration of CrMnFeCoNi HEA was prepared by randomly assigning element types to each lattice point in the FCC perfect crystal simulation cell with X, Y and Z principal axis aligned with [100], [010] and [001] crystal orientations respectively. Periodic boundary conditions are applied on all directions. The simulation cell contains $30 \times 30 \times 30$ FCC unit cells with overall 108000 constituent atoms. The simulation procedures were performed using Large-scale Atomic/Molecular Massively Parallel Simulator (LAMMPS) package.[66] The force field developed by Lee and coworkers was used to address the interaction among the constituent atoms of CrMnFeCoNi HEA.[67] Energy minimization was performed on the initial configuration based on the conjugate gradient algorithm and the simulation cell was equilibrated under NPT ensemble at 77K with zero external pressure to obtain the quiescent state structure. The constant-volume tensile deformation was applied on the equilibrated simulation cells by changing the boundary condition under constant strain rate $\dot{\gamma}_T = \frac{1}{L_Z}\frac{dL_Z}{dt} = 5 \times 10^8 \ s^{-1}$, where $L_Z$ is the length of simulation box along the Z direction. Simultaneously $L_X$ and $L_Y$, the lengths of simulation box along the X and Y directions respectively, are accordingly adjusted to keep the cell volume and the X-Y aspect ratio invariant. Each timestep of MD simulation was 0.001 $ps$ and trajectory was collected every 1 $ps$ until $\gamma_T$ reached 0.4. The same procedure was repeated on a pure Ni simulation cell for comparison. The lattice constants for CrMnFeCoNi and Ni in our simulations are selected to be 3.598 Å and 3.527 Å respectively based on the experimental results.[9]

**Calculation and analysis of vibrational spectrum.** The vibrational spectrum of HEA was calculated from the particle trajectory generated by MD simulations. The energy minimized structure was first perturbed by randomly assigning velocity to each particle based on a Gaussian distribution with a mean of zero and the standard deviation scaled corresponding to the temperature 0.1 K to properly reflect the local structural evolutions meanwhile ensure the numerical stability. The simulation cell was equilibrated under microcanonical (NVE) ensemble and the particle trajectory was collected for 1 ns with the time interval of 0.1 ps. The monitored region was a 2-layer thick slab perpendicular to the tensile direction with a total of 3600 atoms passing through the initiation site of the first pristine partial dislocation loop. To study the soft mode distribution we calculate the displacement correlation matrix $\langle A_{ij} \rangle$ defined by Mari,[31] where $A_{ij} = r_i(t)r_j(t) - \langle r_i \rangle \langle r_j \rangle$, the index $i$ and $j$ ranging from 1 to 3N, where N is the particle number, and the bracket denotes time average over all monitored time interval. The frequency of each eigenmodes is proportional to the eigenvalue of $A^{-1/2}$ and for each mode the participation fraction of the $i$-th component is then calculated from the corresponding eigenvector: $p_i = |\vec{e}_\omega^i|^2$. By summing the participation fraction over the 5% lowest frequency modes and normalized by dividing by the average value of all monitored particles, $\overline{P_i}$ contributed by each particle to the low-frequency vibrational modes was obtained. Moreover, the particle configuration of the frame right before the emergence of defect avalanche was used as the initial input for vibrational property analysis in the energy minimization process to investigate the spatial distributions of soft spots around the yield point.

**Statistics and excess local average.** In this study the excess local average $\langle \alpha_i \rangle$ of a standardized property $\alpha$ around particle $i$, with respect to the global average of $\alpha$, is defined by convoluting the distribution of $\alpha$ with a Gaussian function whose $\sigma$ is set to be equal the nearest neighbor distance defined by local PDF. The explicit expression of $\langle \alpha_i \rangle$ using particle based discrete values is $\langle \alpha_i \rangle = \frac{\sum_j \exp\left(-\frac{r_{ij}^2}{2\sigma^2}\right)\alpha_j}{\sum_j \exp\left(-\frac{r_{ij}^2}{2\sigma^2}\right)} - \frac{1}{N}\sum_{j=1}^{N} \alpha_j$, where $r_{ij}$ is the distance between particle $i$ and particle $j$ and $N$ is the particle number. The Pearson's correlation coefficient describing the correlations between two properties $\alpha$ and $\beta$ was defined by $r_{\alpha\beta} = \frac{\sum_{i=1}^{N}(\alpha_i - \bar{\alpha})(\beta_i - \bar{\beta})}{\sqrt{\sum_{i=1}^{N}(\alpha_i - \bar{\alpha})^2}\sqrt{\sum_{i=1}^{N}(\beta_i - \bar{\beta})^2}}$, where $N$ is the particle number within the monitored region and $\bar{\alpha}$ is the average of $\alpha$ among the $N$ particles.

**Trajectory analysis.** From the stress-strain relation we focus on the point when the von-Mises stress begins to drop. Around sharp turning point using the common neighbor analysis (CNA) at a certain frame formation of numerous crystal defects was identified and was defined as yield point in this study. Local non-affine displacement around the $i^{th}$ particle was obtained by determining the minimized sum among the first nearest neighbor of the $i^{th}$ particle[18]: $D_i^2 = \sum_n (\Delta r_n(t) - \Delta r_n(0) \cdot J_i)$ where $\Delta r_n(t) = r_n(t) - r_i(t)$ is the displacement from the center particle and $J_i$ is the local deformation gradient tensor around the $i^{th}$ particle. Taking the perfect crustal FCC as the reference structure, the $D_i^2$ of the energy minimized structure is also calculated to quantify the local inherent structure distortion, denoted as $\delta$ in the main text. The time frame of the quiescent state was selected as the reference point. In addition, the local shear strain was also calculated[36] from the invariant $\gamma_l = \sqrt{\eta_{yz}^2 + \eta_{xz}^2 + \eta_{xy}^2 + \frac{(\eta_{yy}-\eta_{zz})^2 + (\eta_{xx}-\eta_{zz})^2 + (\eta_{xx}-\eta_{yy})^2}{6}}$ where $\eta_i = \frac{1}{2}(J_i J_i^T - I)$ is the Lagrangian strain tensor. The identification and visualization of dislocation core structure and Burgers vector in the FCC crystal structure was implemented by the dislocation analysis (DXA) algorithm in the OVITO visualization software.[56,68] The trial circuit length and circuit stretchability were properly selected to avoid the misjudgments around dislocation junctions and reliably identify all the common dislocation types in FCC crystal structure.

**PEL sampling by ART method.** The ART algorithm is employed in probing the activation energy spectra for atomic rearrangements in soft spots and stiff zones, respectively. Specifically, random perturbations are applied to the selected atoms and their nearest neighbors with a total displacement of 1.0 Å. When the lowest eigenvalue (*i.e.* curvature) of the PEL is found to be smaller than -0.01 eV Å$^{-2}$, the lanczos algorithm is then applied to ensure convergence of the system to the saddle point with a force tolerance of 0.005 eV. For each central atom, 10 ART searches with different perturbation directions are applied. Finally, statistical histograms are obtained by removing the repeated and failed searches.

**Dislocation mobility simulation.** We generated an FCC simulation cell with X, Y and Z axis aligning with $[1\bar{1}0]$, $[11\bar{2}]$, and $[111]$ crystal directions and applied periodic boundary condition along Y axis. The dimension of simulation cell was 300 Å × 66 Å × 60 Å containing 30 (111) layers along the Z axis. We define three slabs each containing ten (111) layers as the top, medium and bottom slab separately. As an example, an edge dislocation was generated with its core located at the center slab and dislocation line aligned with the Y axis by introducing a local strain field correspond to an $\frac{1}{2}\langle 110\rangle\{111\}$ type edge dislocation following the theory[26,69] for Volterra's dislocations in elastic medium. After relaxing the particles within the center slab with all the other atoms in the upper and lower slabs fixed, the edge dislocation dissociated into an extended dislocation containing a pair of $\frac{1}{6}\langle 112\rangle\{111\}$ Shockley partial. Next, we applied a XZ shear strain with a strain rate $\dot{\gamma}_T$ of $1 \times 10^8$ s$^{-1}$ to the center slab by controlling the boundary condition similar to the approach proposed by Rodney[70] to drive the dislocation motion. The top slab was displaced along the X axis at a constant velocity relative to the bottom slab. During the deformation process the local stress accumulated and once the Peierls barrier was overcame the dislocations started to slip along the X axis. The particle trajectory as well as the dislocation core coordinate was recorded every 0.1 ps. To investigate the mobility of the leading partial, we assigned a 2D function defined by $f(\boldsymbol{r}, t) = H(\boldsymbol{v}_i(\boldsymbol{r}) \times \boldsymbol{\xi}_{i,t} \cdot \hat{\boldsymbol{z}})$. Where $\boldsymbol{v}_i$ is the vector form the coordinate $\boldsymbol{r}$ on the slip plane to a point $i$ on the dislocation core at time $t$ with minimum length, $\boldsymbol{\xi}_{i,t}$ is the dislocation line vector on point $i$ and $\hat{\boldsymbol{z}}$ is the normal of dislocation slip plane. $H(x) = \frac{1}{2}(1 + \text{sgn}(x))$ is the Heaviside step function. The integration $I(\boldsymbol{r}) = \int_0^T f(\boldsymbol{r}, t)dt$ with respect to time described the marching of dislocation line on the slip plane until time $T$ and thus the 2D distribution of dislocation speed can be calculated by the reciprocal of gradient of $I$: $v(\boldsymbol{r}) = \frac{1}{|\nabla I(\boldsymbol{r})|}$.

**Data availability**

All relevant data are available from the authors.

**Code availability**

All relevant codes are available from the authors.

**Acknowledgements**

This research was supported by the Laboratory Directed Research and Development Program of Oak Ridge National Laboratory, managed by UT-Battelle, LLC, for the U. S. Department of Energy, and performed at SNS, which is DOE Office of Science User Facilities operated by the Oak Ridge National Laboratory. C.-H. T. thanks the financial support from the Shull-Wallen Center during his visit of Oak Ridge National Laboratory.

**Author contributions**

C.-H. T. performed the MD simulation and trajectory analysis. S.-Y. C. designed the experiment. W.-R. C. design methodologies of trajectory analysis and provided guidance. Y. F. and Z. B. conducted the PEL calculations. G.-R. H. provided theoretical suggestions on the calculation of vibrational properties